\documentstyle[12pt]{article}
  
\catcode`@=11
\def\secteqno{\@addtoreset{equation}{section}%
\def\theequation{\thesection.\arabic{equation}}}
\catcode`@=12
\secteqno
\newcommand{\be}{\begin{equation}}
\newcommand{\ee}{\end{equation}}
\newcommand{\bea}{\begin{eqnarray}}
\newcommand{\eea}{\end{eqnarray}}
\newcommand{\bref}[1]{(\ref{#1})}
\newcommand{\nn}{\nonumber}

\newcommand{\mapright}[1]{\smash{[mathop{\hbox to 1cm{\rightarrowfill}}
\limits^{#1}}}

\newcommand{\dfer}{D}
\begin{document}

\begin{flushright}
\parbox{4.2cm}
{KEK-TH-1486 \hfill \\
KUNS-2354    \\
 }
\end{flushright}

\vspace*{1.1cm}

\begin{center}
 \Large\bf Super Yangian of superstring on AdS$_5$ $\times$ S$^5$  revisited
\end{center}
\vspace*{1.5cm}
\centerline{\large Machiko Hatsuda$^{\dagger\ast a}$ and Kentaroh
Yoshida$^{\ast b}$}
\begin{center}
$^{\dagger}$\emph{Physics Department, Juntendo University, 270-1695, Japan}
\vspace*{0.5cm}
\\
$^{\ast}$\emph{Theory Division, High Energy Accelerator Research 
Organization (KEK),\\
Tsukuba, Ibaraki 305-0801, Japan.} 
\vspace*{0.5cm}
\\$^{\dagger}$\emph{Department of Physics, Kyoto University
Kyoto 606-8502, Japan.} 
\vspace*{1cm}
\\
$^{a}$mhatsuda@sakura.juntendo.ac.jp
~~~~
 $^{b}$kyoshida@gauge.scphys.kyoto-u.ac.jp
\end{center}

\vspace*{1cm}

\centerline{\bf Abstract}
  
\vspace*{0.5cm}
We construct infinite number of conserved nonlocal charges 
for type IIB superstring on the AdS$_5\times$S$^5$ space
in the conformal gauge without assuming any $\kappa$ gauge fixing,
and show that they satisfy the super Yangian algebra.
The resultant algebra is the same as our previous work
  \cite{Hatsuda:2004it},
  where a special gauge was assumed in such a way that 
  the Noether current satisfies a flatness condition. 
However the flatness condition for the Noether current of a superstring
on the AdS space is broken in general.
We show that the anomalous contribution is absorbed into the current
where fermionic constraints play an essential role, and
a resultant conserved nonlocal charge has different expression
satisfying the same super Yangian algebra.
\vfill 

\thispagestyle{empty}
\setcounter{page}{0}
\newpage 
\section{Introduction and summary}

Integrability of the AdS/CFT correspondence \cite{integrability} has a possibility to broaden 
its application range from weak to strong coupling.
Yangian symmetry is a symmetry responsible for 
integrable system
\cite{Yangian,Yangian:review}.
Then Yangian symmetry is widely studied 
 for a superstring on AdS spaces
\cite{BPR,Alday}
as well as spin chain systems \cite{spinchain} 
and  CFT duals \cite{dualsucon}.
 Supersymmetry is one of the guiding principles to 
establish the quantum integrability.
However  it has not been confirmed yet whether 
nonlocal charges for a superstring on AdS spaces 
satisfy the super Yangian algebra, because treatment of fermions 
is still not clear. 
We  presented a classical super Yangian algebra for 
a superstring on the AdS$_5\times$S$^5$ 
in the canonical formulation in 
\cite{Hatsuda:2004it},
where a special gauge was assumed in such a way that 
  the Noether current satisfies a flatness condition. 
The existence of this $\kappa$ gauge is not justified yet,
but this gauge is required for the gauged coset model 
as a consistency condition.
In this work we have reexamined the flatness condition to 
construct conserved nonlocal charges.
Then we evaluate brackets of the nonlocal charges
showing that they satisfy the super Yangian algebra
as same as \cite{Hatsuda:2004it}.

 Our starting point is a superstring action which has the global super-AdS symmetry.
 The global invariance guarantees the existence of the Noether current $J^R_\mu$  
 satisfying $\partial^\mu J^R_\mu=0$.
 The index $R$ stands for right-invariant.
  This Noether current does not satisfy the flatness condition without assuming any $\kappa$ gauge fixing,
\bea
&{\cal D}_\mu=\partial_\mu-2J^R_\mu\,, ~~
 \quad 
[\partial^\mu,{\cal D}_\mu]=0\,,\quad 
\epsilon^{\mu\nu}[{\cal D}_\mu,{\cal D}_\nu]=-4\Xi \,\label{DDXi0}
&\eea
where anomalous term is bilinear of ``fermionic" current $q_\mu$
\footnote{
This ``fermionic" current is not 
fermionic, but it is G-valued  as 
\bea
q_\mu=Z\left(Z^{-1}\partial_\mu Z\right)|_{\rm fermi}Z^{-1}
=\left\{
\begin{array}{l}
q_\tau=-
Z\left(\begin{array}{cc}0&(\bar{j}_\sigma){}_{\bar{b}a}\\
(j_\sigma){}_{b\bar{a}}&0\end{array}\right)Z^{-1} 
\approx 2Z\left(\begin{array}{cc}0&D{}_{a\bar{b}}\\
\bar{D}{}_{\bar{a}b}&0\end{array}\right)Z^{-1}
\\
q_\sigma=Z\left(\begin{array}{cc}0&({j}_\sigma){}_{a\bar{b}}\\
(\bar{j}_{\sigma}){}_{\bar{a}b}&0\end{array}\right)Z^{-1}
\end{array}
\right.\label{qmu} ~~.
\eea
$Z$ is a coset parameter of G/H with
``global super AdS group" G and ``local Lorentz group" H, transforming 
 $Z\to gZh$ with $g \in$ G and $h \in$ H.
In canonical formulation $\tau$ derivative is determined by a bracket
 with the Hamiltonian,
so $\tau$ components of fermionic left invariant currents are
 ${j}_\sigma$ and $\bar{j}_\sigma$ as familiar in a flat case.
We denote $\approx$ for the use of fermionic constraints.
 }
\bea
\Xi=\frac{1}{2}\left[q_\tau,q_\sigma\right]~~.\label{RIcnonflatXi}
\eea
Constrast to \bref{DDXi0} we found a flat current by
adding $q_\mu$ with an imaginary coefficient as
\bea
&\tilde{J}^R_\mu=
J^R_{\mu}+\displaystyle\frac{i}{2}\epsilon_{\mu\nu}q^\nu&~~~\label{qtildet}\nn\\
&\tilde{\cal D}_\mu=\partial_\mu-2\tilde{J}^R_\mu, \,\quad
[\partial^\mu,\tilde{\cal D}_\mu]=4i\Xi,  \, \quad 
\epsilon^{\mu\nu}[\tilde{\cal D}_\mu,\tilde{\cal D}_\nu]=0\, \quad 
&\label{DDXi}
\eea
where its conservation is broken
The question is how to make  conserved nonlocal  charges
 from these two covariant derivatives,
 and whether they satisfy super Yangian algebra.

After deriving these above relations in section 2.1,
we construct a set of infinite number of conserved nonlocal currents
in section 2.2.
In section 3.1 we 
construct the nonlocal charge in the form of the
sum of the Bena-Polchinski-Roiban (BPR) connection 
\cite{BPR}
 and fermionic constraint
in such a way that it commutes 
with the fermionic constraint.
The modification of the BPR connection 
by Hamiltonian constraints 
 is expected in \cite{Magro:2008dv}.
The property that 
the nonlocal charge commutes with the 
fermionic constraints
 is crucial for the practical computation of 
the algebra where the Poisson bracket is allowed to 
use instead of the Dirac bracket.
This also confirms the $\kappa$-symmetry invariance of the super Yangian charges.
In section 3.2 we compute the super Yangian algebra
which is the same as our previous work
with different expression of generators.

\par\vskip 6mm
\section{Super Yangian generators}

In this section we construct nonlocal charges of the
AdS$_5\times$S$^5$ superstring as  super Yangian generators.
At first we derive several current relations such as flatness 
and conservation in the conformal gauge without assuming any other gauge fixing.
Using these relations we construct conserved nonlocal currents. 
 
\par\vskip 6mm
\subsection{Flat currents for AdS$_5\times$S$^5$ superstring}

The notation follows from \cite{Hatsuda:2004it}.
We use the Roiban-Siegel action for a  superstring on AdS$_5\times$S$^5$ 
\cite{RS} which is based on
a coset G/H with G=GL(4$\mid$4) and H=[Sp(4)GL(1)]$^2$.
A coset parameter $Z_M{}^{A}$ which is transformed as $Z\to gZh$  with $g\in$ G and $h\in$ H.
Left-invariant (LI) currents are denoted by
\bea
(J_\sigma^{L})_A{}^B=(Z^{-1})_A{}^M\partial_\sigma Z_M{}^B=\left(\begin{array}{cc}
{\bf J}_\sigma&j_\sigma\\
\bar{j}_\sigma&\bar{\bf J}_{\sigma}
\end{array}\right)~~,\label{LIJs}
\eea
where notation of components of supermatrices are in  footnote
\footnote{
A supermatrix is denoted by boldfaced letters  for bosonic components
and small letters for fermionic components as
\bea
M_{AB}=\left(\begin{array}{cc}
{\bf M}_{ab}&m_{a\bar{b}}\\\bar{m}_{\bar{a}b}&\bar{\bf M}_{\bar{a}\bar{b}}
\end{array}\right)~~,~~
{\bf M}_{ab}=({\bf M})_{(ab)}+\langle {\bf M}\rangle_{\langle ab\rangle}+
\frac{1}{4}\Omega_{ab}{\rm tr}{\bf  M}
\eea
with  symmetric part $(ab)$,  
traceless-antisymmetric part $\langle ab\rangle$, 
and trace part ${\rm tr}{\bf M}=\Omega^{ab}{\bf M}_{ab}
$ respectively. $\Omega_{AB}$ is antisymmetric Sp(4)$^2$ invariant metric.
}.
The canonical conjugate to $Z_M{}^A$ is $\Pi_A{}^M$
satisfying  
$
[Z_M{}^A,\Pi_B{}^N\}_{\rm P}=(-)^A\delta^A_B\delta^N_M
$.
The bracket is the  graded Poisson bracket
$[A,B\}_{\rm P}=\frac{\partial A}{\partial Z}
\frac{\partial B}{\partial \Pi}
-(-)^{\sigma(Z)}
\frac{\partial A}{\partial \Pi}
\frac{\partial B}{\partial Z}
$, 
 and should not be confused with 
the commutator of matrices $[A,B]=AB-BA$.
The LI supercovariant derivative is
given as
\bea
D_A{}^B=\Pi_{A}{}^MZ_M{}^B=
\left(\begin{array}{cc}
{\bf D}&\dfer \\\bar{\dfer}&\bar{\bf D}
\end{array}\right)~~~.\label{LID}
\eea

The Hamiltonian of the system in the conformal gauge is given by \cite{HK}
\bea
{\cal H}&=&-\displaystyle\int ~d\sigma~{\rm tr}
\left[\displaystyle\frac{1}{2}\left\{
\langle{\bf D}\rangle^2+\langle{\bf J}_\sigma\rangle^2+
-\langle\bar{\bf D}\rangle^2-\langle\bar{\bf J}_\sigma\rangle^2\right\}
\right.\nn\\
&&~~~~~~~~~~~~~~~~~~~~~~~\left.
+(\bar{D}\bar{j}_\sigma-Dj_\sigma+j_\sigma \bar{j}_\sigma)\right]~~~.
\label{Hamiltonian}
\eea
We use full GL(4$\mid$4) parameters $Z_M{}^A$ by gauging H 
components, so $Z_M{}^A$ is constrained by H-gauge symmetry.
In addition fermionic constraints exist whose half generate
the $\kappa$-symmetry. 
H-gauge constraints and fermionic constraints are 
\bea
&({\bf D})_{(ab)}={\rm tr}{\bf D}=(\bar{\bf D})_{(\bar{a}\bar{b})}={\rm tr}\bar{\bf D}=0&\nn\\
&F_{a\bar{b}}=E^{1/4}D_{a\bar{b}}+\frac{1}{2}E^{-1/4}(\bar{j}_\sigma){}_{\bar{b}a}=0&\nn\\
&\bar{F}_{\bar{a}{b}}=E^{-1/4}\bar{D}_{\bar{a}{b}}+\frac{1}{2}E^{1/4}({j}_\sigma){}_{{b}\bar{a}}=0&\label{constraints}~~~
\eea
with $E={\rm sdet} Z$.
Poisson brackets between these constraints and Hamiltonian in \bref{Hamiltonian}
are zero.

Equations of motions  are determined by 
the Poisson bracket  with the Hamiltonian in \bref{Hamiltonian}
as $\partial_\tau {\cal O}=\left[{\cal O} ,{\cal H}\right]$,
\bea
\partial_\tau Z&=&\left[
Z,{\cal H}\right]_{\rm P}~=~
Z\left(
\begin{array}{cc}
\langle {\bf D}\rangle &-\bar{j}^T_\sigma\\
-j^T_\sigma&\langle\bar{\bf D}\rangle
\end{array}
\right)~~~.
\eea
The $\tau$ derivative of LI currents 
in \bref{LIJs} and \bref{LID} 
are given as
\bea
\partial_\tau \langle {\bf D}\rangle&=&
\partial_\sigma \langle {\bf J}_\sigma \rangle +
\left[({\bf J}_\sigma),\langle {\bf J}_\sigma\rangle\right]~~,~~
\partial_\tau \langle \bar{\bf D}\rangle~=~
\partial_\sigma \langle \bar{\bf J}_\sigma \rangle +
\left[(\bar{\bf J}_\sigma),\langle \bar{\bf J}_\sigma\rangle\right]\nn\\
\partial_\tau \langle {\bf J}_\sigma\rangle&=&
\partial_\sigma \langle {\bf D}\rangle +
\left[({\bf J}_\sigma),\langle {\bf D}\rangle\right]~~,~~
\partial_\tau \langle \bar{\bf J}_\sigma\rangle~=~
\partial_\sigma \langle \bar{\bf D}\rangle +
\left[(\bar{\bf J}_\sigma),\langle \bar{\bf D}\rangle\right]\nn\\
\partial_\tau j_\sigma/2&=&
\partial_\sigma {\dfer} +{\bf J}_\sigma {\dfer}- {\dfer}\bar{\bf J}_\sigma
+\{j_\sigma\langle \bar{\bf D}\rangle-\langle {\bf D}\rangle j_\sigma\}/2\nn\\
\partial_\tau \bar{j}_\sigma/2&=&
\partial_\sigma \bar{\dfer} +\bar{\bf J}_\sigma \bar{\dfer}- \bar{\dfer}{\bf J}_\sigma
+\{\bar{j}_\sigma\langle {\bf D}\rangle-\langle \bar{\bf D}\rangle \bar{j}_\sigma\}/2~~~.
\label{eomLI}
\eea
In general the right hand side of the first line 
contains bilinear of fermionic currents $\langle j_\sigma
\bar{ {\dfer}}\rangle- \langle{\dfer}\bar{j}_\sigma\rangle$,
however it vanishes in this case 
by fermionic constraint
and its antisymmetric property.
For example 
$
\langle j_\sigma
\bar{\dfer}\rangle \approx 
(j_\sigma){}_{\langle a}{}^{\bar{a}} 
 (j_\sigma){}_{b\rangle}{}^{ \bar{b}} \epsilon_{\bar{b}\bar{a}}
 =0$.

The Noether current, which is right-invariant (RI), 
is given by
\bea
\partial^\mu J^R_\mu=0~~,~~
J^R_\mu=\left\{
\begin{array}{l}
J^R_\tau=Z\Pi=ZDZ^{-1}\\
J^R_\sigma=Z(J_L+{\cal A}) Z^{-1}=
Z\langle J^L_\sigma\rangle Z^{-1}
\end{array}
\right.~~~.
\eea
with 
\bea
\langle J^L_{\sigma}\rangle
 \equiv \left(
\begin{array}{cc}
\langle {\bf J}_\sigma \rangle& 
\frac{1}{2}{j}_\sigma \\ 
\frac{1}{2}\bar{j}_\sigma& 
\langle \bar{\bf J}_\sigma \rangle
\end{array}
\right)~,~\label{kakkoJ}
{\cal A}=
\left(
\begin{array}{cc}
{\bf A}&-
\frac{1}{2}{j}_\sigma \\ -
\frac{1}{2}\bar{j}_\sigma& 
\bar{\bf A}\end{array}
\right)
~,~
\left\{\begin{array}{l}
-{\bf A}=( {\bf J}_\sigma )+
\frac{1}{4}\Omega_{ab}~{\rm tr}{\bf J}_\sigma\\
-{\bf A}=
( \bar{\bf J}_\sigma )+
\frac{1}{4}\Omega_{\bar{a}\bar{b}}~{\rm tr}\bar{\bf J}_\sigma \end{array}\right.
\eea
The bosonic part of
${\cal A}$ is gauge field for gauged H-symmetry of the coset G/H,
while fermionic part of ${\cal A}$ is reflection of the 
fermionic constraint so is not able to gauge away.

In order to calculate the flatness condition of the Noether current,
the following relation is used from \bref{eomLI} and \bref{kakkoJ} as
\bea
\partial_{\tau} \langle J^L_{\sigma}\rangle
&=&\partial_{\sigma} D+[D,{\cal A}]-\left[
\left(\begin{array}{cc}0& {\dfer}\\\bar{ {\dfer}}&0\end{array}\right)
,\langle J^L_{\sigma}\rangle
\right]+\xi 
\eea
\bea
\xi&=&\left[
\left(\begin{array}{cc}0& {\dfer}\\\bar{ {\dfer}}&0\end{array}\right),
\left(\begin{array}{cc}0&j_\sigma\\\bar{j}_{\sigma}&0\end{array}\right)
\right]~~~.\nn
\eea
In the previous paper $\xi$ was absent,
since the fermionic constraints make 
$\xi=\Bigl(\xi_{(ab)},~\xi_{(\bar{a}\bar{b})}\Bigr)$  
to be elements of H which might be gauged away consistently.
In this paper we keep this term 
and recalculate the conserved nonlocal currents and
the super Yangian algebra. 
The flatness condition is broken by the $\xi$ term as
\bea
\partial_{\tau} J^R_{\sigma}-\partial_{\sigma} J^R_{\tau}
-2(J_{\tau}^RJ_{\sigma}^R-J_{\sigma}^RJ_{\tau}^R)=Z\xi Z^{-1}
~~~\label{RIcnonflat}
\eea 
This flatness anomaly is recognized as $\Xi$ in \bref{RIcnonflatXi}
by the use of the fermionic constraints in \bref{constraints},
\bea
Z\xi Z^{-1}=\frac{1}{2}(q_\tau  q_\sigma  - q_\sigma  q_\tau)=\Xi~~~.\nn
\eea
On the other hand we found a modified current in 
\bref{qtildet} which is flat
 \footnote{Our notation is  $\epsilon^{\mu\nu}\epsilon_{\mu\rho}=\delta^\nu_\rho$, 
$\epsilon^{\tau\sigma}=\epsilon_{\tau\sigma}=1$.
Then $\epsilon^{\mu\nu}q_\mu q_\nu=-
\epsilon_{\mu\nu}q^\mu q^\nu
$.
}
\bea
&\partial_{\tau} \tilde{J}^R_{\sigma}-\partial_{\sigma} \tilde{J}^R_{\tau}-
2(\tilde{J}_{\tau}^R\tilde{J}_{\sigma}^R-\tilde{J}_{\sigma}^R\tilde{J}_{\tau}^R)~=~0
~~&
\eea
but it is not conserved
\bea
\partial^\mu \tilde{J}_\mu^R~=~-2i\Xi~~.\label{dtJX}
\eea
The fact that conservation anomaly $\Xi$ in \bref{dtJX}
is the same function appeared in the flatness anomaly in \bref{RIcnonflat}
leads to another non-trivial flat current 
\bea
\partial_\tau q_\sigma-\partial_\sigma q_\tau
+q_\tau q_\sigma-
q_\sigma q_\tau~=~0
~~~.\label{curlq}
\eea
But it is not conserved
\bea
\partial_\tau q_\tau-\partial_\sigma q_\sigma=
2(J^R_\tau  q_\tau -  q_\tau J^R_\tau - J_\sigma^R
 q_\sigma + q_\sigma J_\sigma^R )~~~.\label{divq}
\eea
The flatness of $q_\mu$ is essential to construct nonlocal currents,
where the flatness anomaly in \bref{RIcnonflat} 
is converted into
divergence of a current as
\bea
\partial_{\tau} (J^R_{\sigma}-\displaystyle\frac{1}{4}
q_\sigma)
-\partial_{\sigma} (J^R_{\tau}
-\displaystyle\frac{1}{4}
q_\tau)
&=&
2(J_{\tau}^R  J_{\sigma}^R
-J_{\sigma}^R  J_{\tau}^R)~~~.\label{J1source}
\eea
This modified flatness condition 
is nothing but the conservation law of  the first level nonlocal current. 
\par \vskip 6mm

\subsection{Conservation of nonlocal currents}

Conserved non-local currents are constructed
quite analogous to the inductive method by 
Brezin, Izykson, Zinn-Justin and Zuber \cite{BIZZ} (BIZZ)
with our non-flat covariant derivative ${\cal D}_\mu$ in
\bref{DDXi0}.
  The $0$-th level of conserved current is the Noether current 
$({\cal J}_0)_\mu=J^R_\mu$.
It can be written as $J^R_\mu=\epsilon_{\mu\nu}\partial^\nu\chi_0$.
Let us set $\chi_{-1}=-\frac{1}{2}$
in such a way that $({\cal J}_0)_\mu={\cal D}_\mu \chi_{-1}$.
According to the BIZZ procudure the $1$-st level conserved current includes
${\cal D}_\mu\chi_0$.
This term is not conserved
\bea
\partial^\mu \left({\cal D}_\mu\chi_0\right)
=-\frac{1}{2}\epsilon^{\mu\nu}\left[
{\cal D}_\mu,{\cal D}_\nu\right] \chi_{-1}
=2\Xi \chi_{-1}
=\frac{1}{4}\partial^\mu~(\epsilon_{\mu\nu} q^\nu)
\nn~~~,
\eea
but the anomalous term is converted into divergence of a current.
The obtained conserved current is
\bea
({\cal J}_1)_\mu(\sigma)&=&{\cal D}_\mu\chi_0+\frac{1}{2}\epsilon_{\mu\nu}q^\nu\chi_{-1}\nn\\
&=&\epsilon_{\mu\nu}
(J^R-\frac{1}{4}q)^\nu(\sigma)-
2J^R_\mu(\sigma)\displaystyle\int_{}^{\sigma}\!\!d\sigma'\, 
(J^R)_\tau(\sigma')\label{J1J1}
\eea\bea
&\Rightarrow \partial^\mu({\cal J}_1)_\mu=0&~~~\nn
\nn
\eea
where
$\chi_0(\sigma)=\displaystyle\int^\sigma d\sigma'~J^R{}_\tau(\sigma')$
is used. The integration path, denoted by $\displaystyle\int$  and $\displaystyle\int^\sigma$, 
must be chosen to make well defined functions
where a cut in a closed string worldsheet is required \cite{Hatsuda:2004it,Hatsuda:2006ts}.

The second level conserved current includes
${\cal D}_\mu\chi_1$ with $({\cal J}_1)_\mu=\epsilon_{\mu\nu}\partial^\nu\chi_1$,
which is not conserved
\bea
\partial^\mu \left( {\cal D}_\mu\chi_1\right)
&=&\partial^\mu\left(
-\frac{1}{2}\epsilon_{\mu\nu}q^\nu\chi_{0}\right)
~~~.\nn
\eea
The conserved current is obtained as
\bea
({\cal J}_2)_{\mu}(\sigma) 
&=&{\cal D}_\mu\chi_1+\frac{1}{2}\epsilon_{\mu\nu}q^\nu \chi_{0}\nn\\
&=& (J^R-\frac{1}{4}q)_{\mu} (\sigma)\label{J2J2} \\
&&
- 2 \epsilon_{\mu\nu}
 (J^R-\frac{1}{4}q)^{\nu} (\sigma)
\displaystyle\int^{\sigma} d\sigma'\,(J^R)_{\tau}(\sigma') -
 2 J^R_{\mu}(\sigma){\displaystyle\int^{\sigma}}
 d\sigma'\, 
(J^R-{\textstyle \frac{1}{4}}q)_{\sigma}(\sigma')
\nn\\&&+ 4 J_{\mu}^R(\sigma)
\displaystyle\int^{\sigma} d\sigma'\,(J^R)_{\tau}(\sigma')
\displaystyle\int^{\sigma'}
d\sigma''\,(J^R)_{\tau}(\sigma'')\nn
\eea
\bea
&\Rightarrow 
\partial^\mu({\cal J}_2)_\mu=0&~\nn
\eea
It is straightforward to check 
 $\partial_\tau~\displaystyle\int ({\cal J}_2)_\tau=0$ directly
by \bref{divq} and \bref{J1source}.
 
In induction there exists a potential $\chi_n$ for a conserved current, 
$\partial^\mu({\cal J}_n)_\mu=0$, 
\bea
&({\cal J}_n)_\mu=\epsilon_{\mu\nu}\partial^\nu\chi_n  \quad 
(n\geq 0)\,  &
\eea
with $\partial^\mu \chi_n=-\epsilon^{\mu\nu}({\cal J}_n)_\nu$. Acting  ${\cal D}_\mu$  
on $\chi_n$ and converting an anomalous term into a divergence of current
give an infinite number of conserved currents as $\partial^\mu({\cal J}_n)_\mu=0$. 
Conserved currents can be constructed as
\bea
({\cal J}_{n+1})_\mu={\cal D}_\mu \chi_n
+\frac{1}{2}\epsilon_{\mu\nu}q^\nu
\displaystyle\sum_{l=0}^{[n/2]} a_{n-1-2l}\chi_{n-1-2l}
\,,\label{newnonlocal}
\eea
with
$a_{n-1}=1,~a_{n-3}=1/4,~a_{n-5}=1/8,~a_{n-7}=5/64,\cdots~
$ and $a_{n-1-2l}$'s are determined perturbatively.
The obtained conserved nonlocal currents are given by
\bea
\left\{\begin{array}{rcl}
({\cal J}_0)_\mu(\sigma)&=&J^R_\mu(\sigma)\\
({\cal J}_1)_\mu(\sigma)&=&\epsilon_{\mu\nu}
(J^R-\frac{1}{4}q)^\nu(\sigma)+
2J^R_\mu(\sigma)\displaystyle\int_{}^{\sigma}\!\!d\sigma'\, 
(J^R)^\tau(\sigma')\\
({\cal J}_2)_{\mu}(\sigma) &=& (J^R-\frac{1}{4}q)_{\mu} 
+ 2 \epsilon_{\mu\nu}
 (J^R-\frac{1}{4}q)^{\nu} 
\displaystyle\int^{\sigma} d\sigma'\,(J^R)^{\tau}(\sigma')  \\
&&- 2 J^R_{\mu}{\displaystyle\int^{\sigma}}
 d\sigma'\, 
(J^R-{\textstyle \frac{1}{4}}q)_{\sigma}(\sigma')+ 4 J_{\mu}^R
\displaystyle\int^{\sigma} d\sigma'\,(J^R)_{\tau}(\sigma')
\displaystyle\int
d\sigma''\,(J^R)_{\tau}(\sigma'')
\\
\vdots \end{array}\right. . \label{calJ}
 \eea 
There exists infinite number of the conserved nonlocal charges 
$Q_n=\displaystyle\int\!d\sigma ({\cal J}_n)_\tau$.
There is ambiguity of functions of $Q_0$ so we
begin with 
\begin{eqnarray}
 Q_{1} &=& {\displaystyle\int}~
d\sigma~({\cal J}_1)_\tau 
\equiv {\displaystyle\int}~
d\sigma\,(J^R_\sigma-{\textstyle \frac{1}{4}}q_\sigma) (\sigma) 
-{\displaystyle\int}~
d\sigma \displaystyle\int^{\sigma}  d\sigma'\, 
\left[J^R_\tau(\sigma),\, J^R_\tau(\sigma')\right]\nn\\
&=&{\displaystyle\int}~ d\sigma\,
(J^R_\sigma-{\textstyle  \frac{1}{4}}q_\sigma) (\sigma) 
-\frac{1}{2}{\displaystyle\int}~
d\sigma \displaystyle\int~ d\sigma'\,
\epsilon(\sigma - \sigma')
\left[J^R_\tau(\sigma),\, J^R_\tau(\sigma')\right]\,, \label{Q1Q1}
\end{eqnarray}
with $\epsilon(\sigma-\sigma')=\theta(\sigma-\sigma')-\theta(\sigma'-\sigma)$.

It is noted  that our Noether current and the first level nonlocal charge 
are equal to ones obtained by Bena, Polchinski and Roiban
\cite{BPR}
with use of constraints. \footnote{Correspondence with their notation is  the following;
For example Noether current in their notaion is given by
\bea
\left(p+\frac{1}{2}\* q'\right)_\mu=
\left\{\begin{array}{l}
Z
\left(\begin{array}{cc}
(J_\tau){}_{\langle ab\rangle}&-\frac{1}{2}({J}_\sigma){}_{\bar{b}a}\\
-\frac{1}{2}(J_\sigma){}_{b\bar{a}}&(J_\tau)_{\langle\bar{a}\bar{b}\rangle}
\end{array}
\right)
Z^{-1} \\
Z
\left(\begin{array}{cc}
(J_\sigma){}_{\langle ab\rangle}&-\frac{1}{2}({J}_\tau){}_{\bar{b}a}\\
-\frac{1}{2}(J_\tau){}_{b\bar{a}}&(J_\sigma)_{\langle\bar{a}\bar{b}\rangle}
\end{array}
\right)
Z^{-1}
\end{array}\right.\nn~~
\Leftrightarrow~~J^R_\mu=
\left\{\begin{array}{l}
ZDZ^{-1}=Z\Pi \\Z
\left(\begin{array}{cc}
\langle {\bf J}_\sigma\rangle{}_{\langle ab\rangle}&\frac{1}{2}j_\sigma
{}_{a\bar{b}}\\
\frac{1}{2}\bar{j}_\sigma{}_{\bar{a}b}&\langle\bar{\bf J}_\sigma\rangle_{\langle\bar{a}\bar{b}\rangle}\end{array}
\right)Z^{-1}
\end{array}\right.\nn~~~.
\eea
In our notation $\tau$-derivative is determined by \bref{eomLI}
as
\bea
(J_\tau){}_{\langle ab\rangle}={\bf D}_{ab}~~,
~~(J_\tau)_{\langle \bar{a}\bar{b}\rangle}=\bar{\bf D}_{\bar{a}\bar{b}}~~,~~
({J}_\sigma){}_{\bar{b}a}=-j_\sigma{}_{a\bar{b}}=2D_{a\bar{b}}~~,~~
(J_\sigma){}_{b\bar{a}}=-\bar{j}_\sigma{}_{\bar{a}b}=2\bar{D}_{\bar{a}b}\nn
\eea
with use of H-gauge constraints and fermionic constraints in \bref{constraints}.
}
It is unclear whether all other nonlocal charges coincide.

\par
\vskip 6mm
\section{Super Yangian algebra}

In this section we compute classical algebra among nonlocal charges obtained as super Yangian generators in the previous section.
The Green-Schwarz type superstring has fermionic second class constraints
which forces to use the Dirac bracket for the algebra computation.
For an operator which commute with the second class constraints
its Dirac bracket with any operator  
reduces to its Poisson bracket.
At first we will find a nonlocal charge
in such a way that it commutes with
the fermionic constraints. 
Then algebra is calculated by the
Poisson bracket.
The gauge invariance of these generators is also confirmed
as expected.

\par
\vskip 6mm
\subsection{Invariance of super Yangian generators}
Let us examine invariance of super Yangian generators:
\bea
Q_0&=&\displaystyle\int
 d\sigma ~
J^R_\tau
(\sigma)\label{Q0Noether}\\
Q_1&=&{\displaystyle\int
}
~ d\sigma\,
(J^R_\sigma-{\textstyle  \frac{1}{4}}q_\sigma) (\sigma) 
-\frac{1}{2}{\displaystyle\int
}
~d\sigma \displaystyle\int
~ d\sigma'\,
\epsilon(\sigma - \sigma')
\left[J^R_\tau(\sigma),J^R_\tau
(\sigma')\right]~~~.\nn
\eea

At first let us confirm the H-gauge invariance of
the super Yangian charges.
The H-gauge constraints in the first line of \bref{constraints}
generating two Sp(4)'s and two GL(1)'s transformations 
are 
\bea
\phi_i=\Bigl\{
({\bf D}_{(ab)}),~
(\bar{\bf D}_{(\bar{a}\bar{b})}),~
{\rm tr}{\bf D},~{\rm tr}\bar{\bf D}\Bigr\}~~~.
\eea
It is easy to confirm that H-invariance of $Q$'s
\bea
\left[Q_0,\phi_i\right\}_{\rm P}=
\left[Q_1,\phi_i\right\}_{\rm P}
=0~~~.
\eea

Next  let us examine the fermionic constraints in 
the second and third lines of \bref{constraints}
whose  half is first class generating the $\kappa$-symmetry
 and another half is second class.
It is obvious that $[Q_0,F\}_{\rm P}=[Q_0,\bar{F}\}_{\rm P}=0$
since $F,\bar{F}$ are made of LI currents.
Then the Dirac bracket between $Q_0$ with any operator ${\cal O}$
is equal to its Poisson bracket, 
 $[Q_0,{\cal O}\}_{\rm Dirac}=[Q_0,{\cal O}\}_{\rm P}$.
 
However the $F$-invariance of $Q_1$ is not realized by itself.
It turns out that fermionic constraints must be added 
to the nonlocal charge $Q_1$ 
in such a way that  a Dirac bracket of $\hat{Q}_1$ with any operator 
are equal to its Poisson bracket as
 as
\bea
\hat{Q}_1&=&Q_1+\int
~d\sigma~ 
Z\left(
\begin{array}{cc}
&\bar{F}^T\\F^T&
\end{array}
\right)Z^{-1}(\sigma)\nn\\
&=&\int
~d\sigma~ 
\biggl\{
Z'Z^{-1}+
Z\left(
\begin{array}{cc}{\bf A}
&-\frac{1}{4}j+\bar{D}^T\\
-\frac{1}{4}\bar{j}+D^T&\bar{\bf A}
\end{array}
\right)Z^{-1}(\sigma)\biggr\}\nn\\
&&-\frac{1}{2}{\displaystyle\int
}~d\sigma \displaystyle\int
~ d\sigma'\,
\epsilon(\sigma - \sigma')
\left[J^R_\tau(\sigma),J^R_\tau(\sigma')\right]~~~.\label{Q1hat}
\eea
\bea
\Rightarrow~\left[
\hat{Q}_1,F\right\}_{\rm P}=\left[
\hat{Q}_1,\bar{F}\right\}_{\rm P}=0~~
\Rightarrow~
[\hat{Q}_1\,, {\mathcal{O}} \}_{\rm Dirac} = [\hat{Q}_1\,, {\mathcal{O}} \}_{\rm P}
~.\nn
&
\eea 

\par
\vskip 6mm
\subsection{Super Yangian algebra}

Now let us calculate the super Yangian algebra.
From now on we denote  $\hat{Q}_1$ by $Q_1$
for simpler notation though, 
fermionic constraints in \bref{Q1hat} 
must be taken into account 
for evaluation of  brackets.

We obtain the classical super Yangian algebra:  
\bea
\left[Q_0{}_M{}^N,
Q_0{}_L{}^K\right\}_{\rm P}&=&
(-)^N\left[s\delta_M^K Q_0{}_L{}^N
-\delta_L^N Q_0{}_M{}^K\right]~~\nn\\
\left[Q_0{}_M{}^N,
Q_1{}_L{}^K\right\}_{\rm P}&=&
(-)^N\left[s\delta_M^K Q_1{}_L{}^N
-\delta_L^N Q_1{}_M{}^K\right]~~\label{Q0Q1}\\
\left[Q_1{}_M{}^N,Q_1{}_L{}^K\right\}_{\rm P}&=&
(-)^N
\Bigl[
s\delta_M^K Q_2{}_L{}^N
-\delta_L^N Q_2{}_M{}^K
\nn\\&&
+4s\left(
Q_0{}_L{}^P Q_0{}_P{}^N Q_0{}_M{}^K
-Q_0{}_L{}^N Q_0{}_M{}^{OP} Q_0{}_P{}^K
\right)
\Bigr]\nn
\eea
where 
\bea
Q_2{}_M{}^N
&=&3Q_0{}_M{}^N+\int ({\cal J}_2)_\tau{}_M{}^N
\label{3}
\eea
with Grassmann sign factor $s=(-)^{(N+L)(1+M+L)}$. 
The resultant algebra is the same as \cite{Hatsuda:2004it}
  but expression of the charge in \bref{3} and \bref{calJ}
is different. 
Details of the computation are given in the appendix.

The Serre relation is followed from \bref{Q0Q1},
so we showed that the nonlocal charge in \bref{Q1hat}
together with the Noether charge in \bref{Q0Noether}
satisfy the super Yangian algebra.  

\par\vskip 6mm
\subsection*{Acknowledgments}
We would like to thank Nathan Berkovits and Dmitri Sorokin
for illuminating discussions at Simons Workshop in Mathematics and Physics. 
The work of KY was supported by the scientific grants from the Ministry of Education, Culture, Sports, Science 
and Technology (MEXT) of Japan (No.\,22740160). 
This work was also supported in part by the Grant-in-Aid 
for the Global COE Program ``The Next Generation of Physics, Spun 
from Universality and Emergence'' from 
MEXT, Japan.

\appendix

\section*{Appendix}

\section{Nonlocal currents} 
The $2$-nd level conserved current includes
${\cal D}_\mu\chi_1$ with $({\cal J}_1)_\mu=\epsilon_{\mu\nu}\partial^\nu\chi_1$
which is not conserved
\bea
\partial^\mu \left( {\cal D}_\mu\chi_1\right)
&=&-\epsilon^{\mu\nu}{\cal D}_\mu
\left({\cal D}_\nu \chi_{0}
-i\Delta {J}_\nu\chi_{-1}
\right)
\nn\\
&=&2\Xi \chi_{0}+i\epsilon^{\mu\nu}{\cal D}_\mu
\left(\Delta {J}_\nu\chi_{-1}\right)
\nn\\
&=&\frac{2}{4i}[\partial^\mu, \tilde{\cal D}_\mu]\chi_{0} 
-i\epsilon^{\mu\nu}
\Delta {J}_\mu {\cal D}_\nu \chi_{-1}
\nn\\
&=&\frac{2}{4i}\left\{\partial^\mu\left(-2\Delta {J}_\mu\chi_{0}\right)
+2\Delta {J}_\mu\partial^\mu\chi_{0}\right\}
-i\epsilon^{\mu\nu}\Delta {J}_\mu (J_0)_\nu
\nn\\
&=&\frac{2}{4i}\left\{\partial^\mu\left(-2\Delta {J}_\mu\chi_{0}\right)
-2\epsilon^{\mu\nu}\Delta {J}_\mu J^R_\nu 
\right\}
-i\epsilon^{\mu\nu}\Delta {J}_\mu J^R_\nu
\nn\\
&=&i\partial^\mu\left(\Delta {J}_\mu\chi_{0}\right)
~~~.\nn
\eea
It is denoted by
$\Delta {J}_\mu=\frac{i}{2}\epsilon_{\mu\nu}q^\nu$.

In induction there exists a potential $\chi_n$
for a conserved current, $\partial^\mu({\cal J}_n)_\mu=0$, 
\bea
&({\cal J}_n)_\mu=\epsilon_{\mu\nu}\partial^\nu\chi_n  \quad 
(n\geq 0)\,  &
\eea
with $\partial^\mu \chi_n=-\epsilon^{\mu\nu}({\cal J}_n)_\nu$
because of notation $\epsilon^{\mu\nu}\epsilon_{\mu\rho}=\delta^\nu_\rho$.
Acting  ${\cal D}_\mu$  
on $\chi_n$ and canceling the anomalies by the anomalous term $\Delta J_\mu$
as \bref{DDXi}
give an infinite number of conserved currents as $\partial^\mu({\cal J}_n)_\mu=0$ 
\bea
({\cal J}_3)_\mu&=&{\cal D}_\mu\chi_2
-i\Delta {J}_\mu(\chi_1+\frac{1}{4}\chi_{-1})\nn\\
({\cal J}_4)_\mu&=&{\cal D}_\mu\chi_3
-i\Delta {J}_\mu(\chi_2+\frac{1}{4}\chi_0)\nn\\
({\cal J}_5)_\mu&=&{\cal D}_\mu\chi_4
-i\Delta {J}_\mu(\chi_3+\frac{1}{4}\chi_1+\frac{1}{8}\chi_{-1})\nn\\
({\cal J}_6)_\mu&=&{\cal D}_\mu\chi_5
-i\Delta {J}_\mu(\chi_4+\frac{1}{4}\chi_2+\frac{1}{8}\chi_0)\nn\\
\cdots~~~.\nn
\eea
In this way conserved currents can be constructed as
\bea
({\cal J}_{n+1})_\mu={\cal D}_\mu \chi_n-i\Delta J_\mu
\displaystyle\sum_{l=0}^{[n/2]} a_{n-1-2l}\chi_{n-1-2l}
\,,\label{newnonlocal}
\eea
with
$a_{n-1}=1,~a_{n-3}=1/4,~a_{n-5}=1/8,~a_{n-7}=5/64,\cdots~
$ and $a_{n-1-2l}$'s are determined perturbatively.
\par
\vskip 6mm
\section{Fermionic constraint invariance of nonlocal charge}

The definition of the Poisson bracket in the footnote 2 
gives convenient formula
\[
\Bigl[\displaystyle\int{\rm str}  \Pi \Psi_1,
\displaystyle\int{\rm str}  Z \Psi_2\Bigr]_{\rm P}=
\displaystyle\int{\rm str}  \Psi_1\Psi_2\,.
\]
In order to compute the Poisson bracket between
the nonlocal charge $Q_1$ in \bref{Q1Q1}
and fermionic constraints $F,\bar{F}$ in \bref{constraints},
we take supertrace with some parameters,
a constant parameter $\Lambda$ for $\hat{Q}_1$ and 
a local parameter $\lambda(\sigma)$ for $F,\bar{F}$,
as 
\bea
\Bigl[{\rm str} \hat{Q}_1 \Lambda
,\displaystyle\int~d\sigma ~{\rm Str}F(\sigma)\lambda(\sigma)\Bigr]_{\rm P}~~,~~~
{\rm Str}F(\sigma)\lambda(\sigma)=
{\rm Str}\left(\begin{array}{cc}
&F(\sigma)\\\bar{F}(\sigma)&
\end{array}\right)\left(\begin{array}{cc}
&\lambda(\sigma)\\\bar{\lambda}(\sigma)&
\end{array}\right)
~~.\nn
\eea
A charge has an ambiguity of the fermionic constraints 
so we examine the following candidate 
\bea
\hat{Q}_1&=&{\displaystyle\int}
~ d\sigma\,
(J^R_\sigma-{\textstyle  \frac{1}{4}}q_\sigma) (\sigma) 
-\frac{1}{2}{\displaystyle\int}
~d\sigma \displaystyle\int~ d\sigma'\,
\epsilon(\sigma - \sigma')
\left[J^R_\tau(\sigma),J^R_\tau
(\sigma')\right]\nn\\
&&+c {\displaystyle\int}
~ d\sigma\,
Z\left(\begin{array}{cc}
&\bar{F}^T\\
F^T&
\end{array}\right)Z^{-1}\nn~~~.
\eea
In appendices we use $\hat{Q}_1$ and ${Q}_1$ separately
in order to stress a role of the fermionic constraints. 
The Poisson bracket between the local term in $\hat{Q}_1$ and $F$ is computed as
\bea
&&\Bigl[{\displaystyle\int}~ d\sigma\,
(J^R_\sigma-{\textstyle  \frac{1}{4}}q_\sigma) (\sigma) 
,\int~d\sigma ~{\rm Str}F\lambda\Bigr]_{\rm P}\nn\\
&&~~~~~~~~~~~~=\int Z\left(
\left[\langle\langle J^L\rangle\rangle,\lambda\right]
-\langle\langle \left[J^L,\lambda\right]\rangle\rangle
-\langle\langle \partial_\sigma\lambda\rangle\rangle
\right)Z^{-1}
\eea
with
\bea
J^R_\sigma-{\textstyle  \displaystyle\frac{1}{4}}q_\sigma&=&Z
\left(\begin{array}{cc}
\langle {\bf J}\rangle& j/4\\
\bar{j}/4&\langle \bar{\bf J}\rangle\end{array}
\right)
Z^{-1}~\equiv~
Z\langle\langle J^L\rangle\rangle Z^{-1}~~~.\nn
\eea
The Poisson bracket of the nonlocal term and $F$ is computed as
\bea
&&\Bigl[-\frac{1}{2}{\displaystyle\int}~
d\sigma \displaystyle\int~ d\sigma'\,
\epsilon(\sigma - \sigma')
\left[
J^R_\tau(\sigma),J^R_\tau
(\sigma')\right]~,
\int~d\sigma ~{\rm Str}F\lambda\Bigr]_{\rm P}\nn\\
&&~~~~~~~~~~~~=
-\int d\sigma ~Z\left[D,\lambda^T\right]Z^{-1}~~,\\
&&~~~~~~~~~~~~\lambda^T=\left(\begin{array}{cc}
&\bar{\lambda}^T(\sigma)\\{\lambda}^T(\sigma)&
\end{array}\right)
\eea
These terms are cancelled 
 by the fermionic constraint as the third term in $\hat{Q}_1$
\bea
&&\Bigl[\displaystyle\int Z\left(
\begin{array}{cc}
&\bar{F}^T\\
F^T&
\end{array}
\right)Z^{-1}, 
\int~d\sigma ~{\rm Str}F\lambda\Bigr]_{\rm P}\nn\\
&&~~~~~~~~~~=Z\left(
\left[\langle{\bf D}\rangle,\lambda^T\right]-\left[\langle{\bf J}\rangle,\lambda\right]
+\left[
\left(
\begin{array}{cc}
&\bar{F}^T\\
F^T&
\end{array}
\right),\lambda
\right]
\right)Z^{-1}~~.
\eea
As a result the Poisson bracket is given by
\bea
&&\Bigl[\hat{Q}_1, 
\int~d\sigma ~{\rm Str}F(\sigma)\lambda(\sigma)\Bigr]_{\rm P}\nn\\
&&~~~~~~~~~~=Z\left(
(c-1)\left[\langle{\bf D}\rangle,\lambda^T\right]
+(1-c)\left[\langle{\bf J}\rangle,\lambda\right]
-\left[
D\mid_{\rm fermi}+\frac{1}{2}J^{L;T}\mid_{\rm fermi}
,\lambda^T
\right]
\right)Z^{-1}~~,\nn\\
&&~~~~~~~~~~=0~~{\rm for} ~~c=1~~~
\eea
where both H-gauge and fermionic constraints are used.
The Dirac bracket of $\hat{Q}_1$ is equal to the Poisson bracket
\bea
[\hat{Q}_1,{\cal O}\}_{\rm Dirac}=[\hat{Q}_1,{\cal O}\}_{\rm P}~~~.
\eea
\par
\vskip 6mm
\section{Derivation of Super Yangian algebra}

Analogous to the previous computation it is convenient to 
multiply parameters as
\bea
\Bigl[{\rm str} ~\hat{Q}_1 \Lambda~
,~{\rm str} ~\hat{Q}_1 \Sigma \Bigr]_{\rm P}~~~.\nn
\eea
The super Yangian generator $\hat{Q}_1$ in \bref{Q1hat}
has H-gauge symmetry which allows a gauge ${\bf A}=\bar{\bf A}=0$ for
simpler computation as
\bea
\hat{Q}_1&=&
\hat{Q}_{1-1}+\hat{Q}_{1-2}\nn\\
\hat{Q}_{1-1}
&=&\int~d\sigma~ 
\biggl\{
Z'Z^{-1}+
Z\left(
\begin{array}{cc}
&-\frac{1}{4}j+\bar{D}^T\\
-\frac{1}{4}\bar{j}+D^T&
\end{array}
\right)Z^{-1}(\sigma)\biggr\}\nn\\
\hat{Q}_{1-2}
&=&-\frac{1}{2}{\displaystyle\int
}~d\sigma \displaystyle\int
~ d\sigma'\,
\epsilon(\sigma - \sigma')
\left[J^R_\tau(\sigma),J^R_\tau(\sigma')\right]~~~.
\eea
The Poisson bracket between two $\hat{Q}_{1-1}$'s is
\bea
\Bigl[{\rm str}~ \hat{Q}_{1-1} \Lambda
,{\rm str}~ \hat{Q}_{1-1} \Sigma \Bigr]_{\rm P}
&=&\displaystyle\int~{\rm str}~
D\mid_{\rm bose}\left[\lambda^T\mid_{\rm fermi},
\sigma^T\mid_{\rm fermi}
\right]\nn
\eea
with $\lambda=Z^{-1}\Lambda Z$
and $\sigma=Z^{-1}\Sigma Z$.
The Poisson bracket between $\hat{Q}_{1-1}$ and
$\hat{Q}_{1-2}$
 is
\bea
&&\Bigl[{\rm str} ~\hat{Q}_{1-1} \Lambda
,{\rm str}~ \hat{Q}_{1-2} \Sigma \Bigr]_{\rm P}
+\Bigl[{\rm str} \hat{Q}_{1-2} \Lambda
,{\rm str} \hat{Q}_{1-1} \Sigma \Bigr]_{\rm P}\nn\\
&&~~~~~~=~\displaystyle\int~{\rm str}~
\Bigl((4J^R_\tau-\frac{1}{4}q_\tau)\left[\Sigma,\Lambda\right]
-D\mid_{\rm bose}\left[\lambda^T\mid_{\rm fermi},
\sigma^T\mid_{\rm fermi}
\right]\Bigr)\nn\\
&&~~~~~~+2\displaystyle\int d\sigma \displaystyle\int d\sigma'
~{\rm str}~
\Bigl[
\left(J_\sigma^R-\frac{1}{4}q_\sigma\right)(\sigma),J_\tau^R(\sigma')
\Bigr]\epsilon(\sigma-\sigma')
\left[\Sigma,\Lambda\right]
\eea
where constraints are set to be zero on the right hand side.
Adding up these terms give 
\bea
&&
\Bigl[{\rm str} ~\hat{Q}_{1-1} \Lambda
,{\rm str} ~\hat{Q}_{1-1} \Sigma \Bigr]_{\rm P}
+\Bigl[{\rm str} \hat{Q}_{1-1} \Lambda
,{\rm str} \hat{Q}_{1-2} \Sigma \Bigr]_{\rm P}
+\Bigl[{\rm str} \hat{Q}_{1-2} \Lambda
,{\rm str} \hat{Q}_{1-1} \Sigma \Bigr]_{\rm P}\nn\\
&&~~~~~~=~\displaystyle\int~{\rm str}~
(4J^R_\tau-\frac{1}{4}q_\tau)\left[\Sigma,\Lambda\right]\nn\\
&&~~~~~~+2\displaystyle\int d\sigma \displaystyle\int  d\sigma'
~{\rm str}~
\Bigl[
\left(J_\sigma^R-\frac{1}{4}q_\sigma\right)(\sigma),J_\tau^R(\sigma')
\Bigr]\epsilon(\sigma-\sigma')
\left[\Sigma,\Lambda\right]~~~.
\eea
The bracket between two $\hat{Q}_{1-2}$'s is the same as our previous result.
Extracting parameters from the above
we get the same final answer as before
\bea
\left[Q_1{}_M{}^N,Q_1{}_L{}^K\right\}_{\rm P}&=&
(-)^N
\Bigl[
s\delta_M^K Q_2{}_L{}^N
-\delta_L^N Q_2{}_M{}^K
\nn\\&&
+4s\left(
Q_0{}_L{}^P Q_0{}_P{}^N Q_0{}_M{}^K
-Q_0{}_L{}^N Q_0{}_M{}^{OP} Q_0{}_P{}^K
\right)
\Bigr]\nn
\eea
where 
\bea
Q_2{}_M{}^N
&=&3Q_0{}_M{}^N+\int ({\cal J}_2)_\tau{}_M{}^N
\nn ~~~.
\eea

\par

\end{document}